\documentclass[aps,prd,nofootinbib,superscriptaddress,preprint,12 pt]{revtex4-1}
\usepackage{amsfonts}
\usepackage{amsmath}
\usepackage{amssymb}
\usepackage{epsfig}
\usepackage{subfigure}
\usepackage{graphicx,color}
\usepackage{booktabs}
\usepackage{multirow}
\usepackage{enumerate}
\usepackage{multirow}
\usepackage[dvipdfm,  
            pdfstartview=FitH,
            CJKbookmarks=true,
            bookmarksnumbered=true,
            bookmarksopen=true,
            linkcolor=blue,
            anchorcolor=blue,
            citecolor=blue
            ]{hyperref}
\usepackage{indentfirst}
\allowdisplaybreaks[1]
\usepackage[dvipsnames]{xcolor}
\begin{document}

\title{Study Standard Model and Majorana Neutrino Contributions to $B^{+} \to K^{(*)\pm}\mu^+\mu^{\mp}$}

\author{Hong-Lei Li}\email{sps$_$lihl@ujn.edu.cn}
\affiliation{School of Physics and Technology, University of Jinan, Jinan Shandong 250022,  China}
\author{Peng-Cheng Lu}\email{pclu@mail.sdu.edu.cn}
\affiliation{School of Physics, Shandong University, Jinan, Shandong 250100,  China}
\author{Cong-Feng Qiao}\email{qiaocf@ucas.ac.cn}
\affiliation{School of Physics, University of Chinese Academy of Sciences, YuQuan Road 19A, Beijing 100049, China}
\affiliation{CAS Center for Excellence in Particle Physics, Beijing 100049, China}
\author{Zong-Guo Si}\email{zgsi@sdu.edu.cn}
\affiliation{School of Physics, Shandong University, Jinan, Shandong 250100,  China}
\author{Ying Wang}\email{wang$_$y@mail.sdu.edu.cn}
\affiliation{School of Physics, University of Chinese Academy of Sciences, YuQuan Road 19A, Beijing 100049, China}
\begin{abstract}
Lepton number violation processes can be induced by the Majorana neutrino exchange, which provide evidence for the Majorana nature of neutrinos. In addition to the natural explanation of the small neutrino masses, Type-I seesaw mechanism predicts the existence of Majorana neutrinos.  The aim of this work is to study the B meson rare decays $B^{+} \to K^{(*)+}\mu^+\mu^-$ in the standard model and its extensions, and then to investigate the same-sign decay processes $B^{+}\to K^{(*)-}\mu^{+}\mu^+$. The corresponding dilepton invariant mass distributions are predicted. It is found that the dilepton angular distributions illustrate the properties of new interactions induced by the Majorana neutrinos.
\\

\end{abstract}

\maketitle
\section{Introduction}
The discovery of the neutrino oscillation \cite{Eguchi:2002dm, Ahmed:2003kj, Argyriades:2008pr, Barger:2003qi, Grinstein:1988me}, confirming the existence of the massive neutrinos, well motivates the search for new physics beyond the SM. Along with the neutrino mass puzzle,  it  is a crucial question to explore the nature of the neutrino. If neutrinos are Dirac particles, lepton number is conserved. Otherwise, lepton number violating (LNV) processes, which are forbidden by SM,  can be induced via a Majorana neutrino exchange.  Therefore   the LNV process will be a promising signal for  new physics beyond the SM. The aim of this paper is to  study the B meson rare decays $B^{+} \to K^{(*)+}\mu^+\mu^-$ in the SM and its extensions, and then to investigate the LNV  process  $B^{+}\to K^{(*)-}\mu^{+}\mu^+$.

Many experimental efforts have been made to search for Majorana neutrinos by $\Delta L=2$ processes, such as the most promising neutrinoless nuclear double beta decays ($0\nu\beta\beta$) \cite{Macolino:2017vyd, Azzolini:2018dyb}, LNV $\tau$ decays \cite{Chrzaszcz:2013uz, Aubert:2005tp} and $\Delta L=2$ $B,D,D_s,K$ meson decays \cite{Aaij:2014aba, Aaij:2013sua, Batley:2011zz}. With the upper limits observed by the experiments, the bounds on the mass of neutrinos and the mixing matrix elements between neutrino and charged leptons are investigated, meanwhile, it can be served to constrain the parameter spaces in the new physics models. Theoretically, for the purpose of proving the Majorana nature of neutrinos, $0\nu\beta\beta$ as well as $\tau^- \to \ell^+ M_1^- M_2^-$ \cite{Helo:2010cw, Castro:2012ma}, $\tau^- \to \pi^+ \mu^- \mu^-  \nu_\tau$ \cite{Dib:2011hc, Castro:2012ma}  and especially many same-sign charged dilepton $B,B_c,B_s,D,D_s,K$ meson decays have been widely investigated, including three-body LNV meson decays $B^+(B_c^+)\to \pi^-(K^{(*)-}, \rho^-, D^{(*)-}, D_s^{(*)-}) \ell^+ \ell^{\prime +} ~(\ell^{(\prime)}=e,\mu)$, $D^+(D_{s}^+)\to \pi^- (K^{(*)-}, \rho^-) \ell^+\ell^{\prime +}$, $K^+\to\pi^- \ell^+\ell^{\prime +}$ \cite{Ali:2001gsa, Atre:2005eb, Atre:2009rg, Cvetic:2010rw, Helo:2010cw, Zhang:2010um, Bao:2012vq, Wang:2014lda, Milanes:2016rzr} and four-body LNV meson decays $\bar{B}^0 \to \pi^+(K^{(*)+}, \rho^+, D_{(s)}^+) D^+ \ell^-\ell^-$, $B^- \to \pi^+ (K^{(*)+}, \rho^+, D_{(s)}^+) D^0 \mu^-\mu^-$, $\bar{D}^0 \to \pi^+(K^+)\pi^+ \mu^-\mu^-$, $B_c^- \to J/\psi \pi^+ \mu^-\mu^-$, $B_c^- \to \bar{B}_s^0  \pi^+ \ell^-\ell^{\prime -}$, $B_s^0\to K^- (D_s^-) \pi^- \mu^+\mu^+$  \cite{Quintero:2011yh, Yuan:2013yba, Dong:2013raa, Castro:2013jsn, Milanes:2016rzr, Mandal:2016hpr, Mejia-Guisao:2017gqp}.
Due to the precision measurements on meson decays, the properties of the Majorana neutrino have been studied on the masses and the mixing matrix elements.

Type-I seesaw mechanism is one of the most natural ways to generate tiny neutrino masses among various new physics models. In this model, the right-handed $\mathrm{SU(2)_{L}\times U(1)_{Y}}$ singlet neutrinos $N_R$ are introduced to extend the SM. Apart from Dirac mass $M_D$, the right-handed neutrino singlets with Majorana mass matrix $M_R$ are allowed with the gauge invariance. As a result, the effective mass matrix for the light neutrinos can be expressed with the canonical seesaw formula $M_{\nu}\thicksim-M_{D}M_{R}^{-1}M_{D}^{T}$.
In terms of the neutrino mass eigenstate, the gauge interaction for the charged current has the formula of
\begin{eqnarray}
-\mathcal{L}=\frac{g}{\sqrt{2}}W_{\mu}^{+}
\Big(\sum_{\ell=e}^{\tau}\sum_{m=1}^{3}U_{\ell m}^{*}\overline{\nu_{m}}\gamma^{\mu}P_{L}\ell
+\sum_{\ell=e}^{\tau}\sum_{m^{'}=4}^{3+n}V_{\ell m^{\prime}}^{*}\overline{N_{m^{\prime}}^{c}}\gamma^{\mu}P_{L}\ell\Big)+h.c.,
\end{eqnarray}
where $P_{L}=(1-\gamma_{5})/2$, $\nu_{m}(m=1,2,3)$ and $N_{m^{\prime}}(m^{\prime}=4,\cdots,3+n)$
are the mass eigenstates, $U_{\ell m}$ ($V_{\ell m^{\prime}}$) is the mixing matrix element between the lepton flavor and light (heavy) neutrinos. Moreover, lots of proposals have been made to search for heavier Majorana neutrinos at $e^- e^-$ ($e^+ e^-$), $e\gamma$, $pp$ ($p\bar{p}$) collider experiments \cite{delAguila:2006bda, Deppisch:2015qwa, Banerjee:2015gca, Bray:2005wv} and also in top quark and W boson rare decays \cite{BarShalom:2006bv, Quintero:2011yh}.

Semileptonic B meson decays $B \to  K^{(*)} \ell^+\ell^-$ induced by flavor changing neutral current are promising processes to test the SM and search for its extension. Precision measurements on the B meson decays have been performed by CDF, BABAR and LHCb collaborations. Recently, the differential branching ratios of $B^{+}\to K^{+}\mu^+\mu^-$ and $B^{+}\to K^{*+}\mu^+\mu^-$ decays have been reported by the LHCb collaboration with the integrated luminosity of $3~\rm{fb}^{-1}$ and the integrated branching fractions are $(4.29\pm0.07(stat) \pm0.21(syst) ) \times10^{-7}$ and $(9.24\pm0.93(stat) \pm0.67(syst) ) \times10^{-7}$ \cite{Aaij:2014pli}, respectively. This is the most precise measurement so far. Along with the development of the experiments, theoretical studies on the $B \to  K^{(*)} \ell^+\ell^-$  have been reported both in the SM \cite{Aliev:1997gh, Ali:1999mm, Khodjamirian:2010vf, Bouchard:2013mia, Fu:2014uea, Ahmady:2015fha, Momeni:2017moz} and new physics models \cite{Crivellin:2015mga, Ahmed:2017vsr, Allanach:2017bta}. These SM predictions for the branching ratios are comparable with the LHCb data, however, new physics contributions can not be excluded. In this paper, we first study the opposite-sign B meson dileptonic rare decays $B^+ \to  K^{(*)+} \mu^+\mu^-$ both in the SM and type-I seesaw model, then the the contributions from Majorana neutrino are investigated in the same-sign LNV processes $B^+ \to  K^{(*)-} \mu^+\mu^+$.

This paper is organized as the follows. In Sec.2, the theoretical framework is introduced with the formulas of B meson rare decays $B^{+}\to K^{(*)+}\ell^{+}\ell^{-}$ in the SM and mediated by Majorana neutrino, as well as the same-sign LNV decays $B^{+}\to K^{(*)-}\ell^{+}\ell^{+}$ .  In Sec.3, we give the numerical results on the branching ratios, the dilepton invariant mass distributions and angular distributions of $B^{+}\to K^{(*)\pm}\mu^{+}\mu^{\mp}$.  The excluded regions of the Majorana neutrino mass and the mixing matrix element are given with the fitting results. Furthermore, the dilepton invariant mass distributions and the dilepton angular distributions of LNV processes are studied. Finally, we give a brief summary.
\section{Theoretical framework}
\begin{figure}[!htb]
\begin{center}
\subfigure[]{ \label{fig:BKlla}
\includegraphics[scale=0.55]{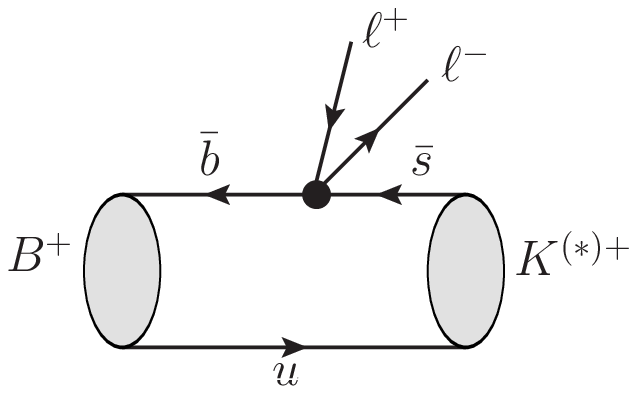}
}
\subfigure[]{ \label{fig:BKllb}
\includegraphics[scale=0.55]{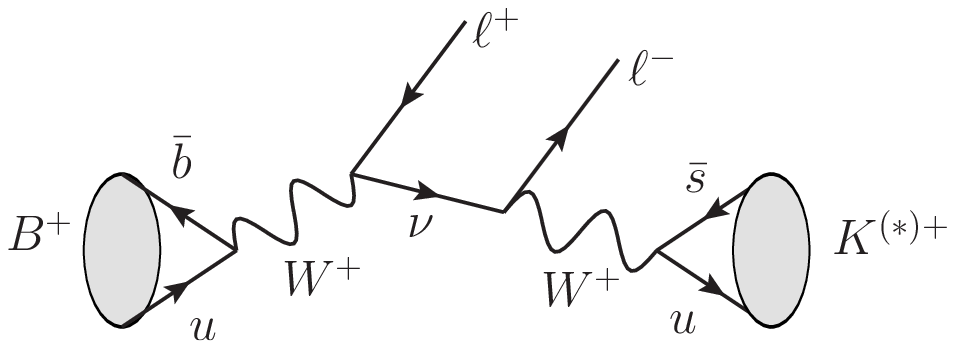}
}
\subfigure[]{ \label{fig:BKllc}
\includegraphics[scale=0.55]{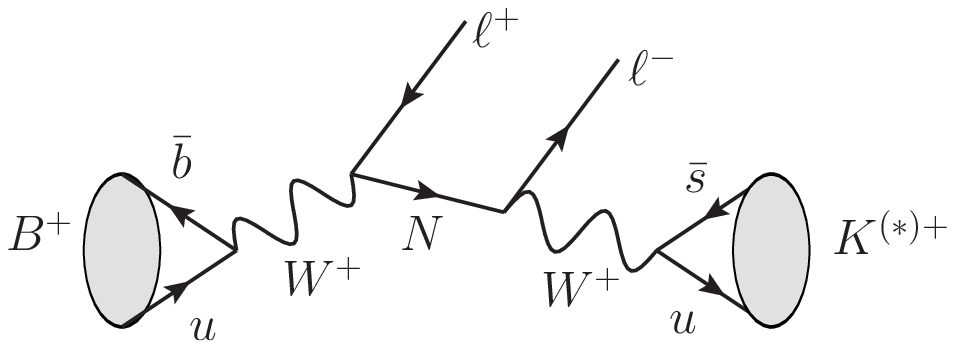}
}
\caption{Feynman diagrams for the opposite-sign dilepton B meson rare decays $B^{+}\to K^{(*)+}\ell^{+}\ell^{-}$. The solid circle stands for the effective vertex of the leading-order $b\to s\ell^+\ell^-$ transition and $N$ denotes the Majorana neutrino.}\label{fig:BKll}
\end{center}
\end{figure}
We first study the B meson rare decays $B^{+}(p_B)\to K^{(*)+}(p_{K^{(*)}}) \ell^{+}(p_1) \ell^{-}(p_2)$. Fig.\ref{fig:BKlla} is the Feynman diagram for $B^{+}\to K^{(*)+}\ell^{+}\ell^{-}$ in the SM. These processes are induced by the flavor changing neutral current $b\to s\ell^+\ell^-$, which can be described through the effective Hamiltonian \cite{Grinstein:1988me}. The decay amplitude of $b\to s\ell^+\ell^-$ can be written as \cite{Ali:1999mm, Ali:2002jg}
\begin{eqnarray}\label{amplitude}
\mathcal{M}(b\to s \ell^+ \ell^-) &=& \frac{G_{F}\alpha}{\sqrt{2}\pi}V^{*}_{ts}V_{tb}\Big\{C_{9}^{eff}[\bar{s}\gamma_{\mu}P_L b][\bar{\ell}\gamma^{\mu}\ell]
+C_{10}[\bar{s}\gamma_{\mu}P_L b][\bar{\ell}\gamma^{\mu}\gamma_{5}\ell]\nonumber\\
&&-2C_{7}^{eff}\left[\bar{s}i\sigma_{\mu\nu}\frac{q^{\nu}}{s}(m_{b}P_{R}+m_{s}P_{L}) b\right][\bar{\ell}\gamma^{\mu}\ell]\Big\}~.
\end{eqnarray}
Here, $P_L=(1-\gamma_{5})/2$, $P_R=(1+\gamma_{5})/2$, $s=q^{2}=(p_1+p_2)^2$.
$G_{F}$ is the Fermi coupling constant. $\alpha$ is the fine-structure constant
and $V_{q_{1}q_{2}}$ is the CKM matrix element. All the Wilson coefficients $C_{i}$ except $C_{9}^{eff}$ have the same analytic expressions as those used in the $b\to s$ transition processes~\cite{Buras:1994dj}, while the $C_{9}^{eff}$ can be found in  \cite{Kruger:1996dt} with the next-to-leading order approximation.
\begin{table}[h]
\centering \caption {The values of Wilson coefficients $C_{i}(\mu)$ in SM with the scale $\mu=m_{b}$
at the leading logarithmic approximation, with $m_{W}=80.4~\mathrm{GeV}$,
$m_{t}=173.5~\mathrm{GeV}$, $m_{b}=4.8~\mathrm{GeV}$ \cite{Wang:2014lda}.}\label{tab:WilsonC}
\begin{tabular}{ccccccccc}
\hline
 $C_{1}$ &$C_{2}$ &$C_{3}$ &$C_{4}$ &$C_{5}$
&$C_{6}$ &$C_{7}^{eff}$  &$C_{9}$ &$C_{10}$\\
\hline
-0.246 &1.106 &0.011 &-0.025 &0.007
&-0.031 &-0.312  &4.211 &-4.501 \\
\hline
\end{tabular}
\end{table}
Generally, exclusive decays $B\to K^{(*)}\ell^{+}\ell^{-}$ are described by matrix elements of the quark operators over meson states, and the matrix elements are ulteriorly be parameterized in terms of $B\to K^{(*)}$ form factors \cite{Ali:1999mm}. For the pseudoscalar $K$ meson, $B\to K$ form factors are defined as follows,
\begin{eqnarray}\label{eq:Matri-Ele-BK}
&&\langle K(p_{K})|\bar{s}\gamma_{\mu}b|B(p_{B})\rangle = (p_{B}+p_{K})_{\mu}f_{+}^{B \to K}(q^2) + \frac{m_{B}^{2}-m_{K}^{2}}{q^{2}}q_{\mu}(f_{0}^{B \to K}(q^2) - f_{+}^{B \to K}(q^2))~,\\
&&\langle K(p_{K})|\bar{s}\sigma_{\mu\nu}q^{\nu}b|B(p_{B})\rangle = i[(p_{B}+p_{K})_{\mu}q^{2}-(m_{B}^{2}-m_{K}^{2})q_{\mu}] \frac{f_{T}^{B \to K}(q^2)}{m_{B}+m_{K}}~.
\end{eqnarray}
Here, $f_{+}^{B \to K}(q^2)$, $f_{0}^{B \to K}(q^2)$ and $f_{T}^{B \to K}(q^2)$ are vector, scalar and tensor form factors of $B\to K$ transition, respectively.
For the vector meson $K^{*}$ with four-momenta $p_{K^*}$ and polarization vector $\epsilon_{\mu}$, the semileptonic form factors of $V-A$ current and the penguin form factors can be defined as
\begin{eqnarray}\label{eq:Matri-Ele-BKst}
&&\langle K^{*}(p_{K^{*}})|\bar{s}\gamma_{\mu}(1-\gamma_{5})b|B(p_{B})\rangle
= -i\epsilon_{\mu}^{*}(m_{B}+m_{K^{*}})A_{1}(q^2)
+i(p_{B}+p_{K^{*}})_{\mu}(\epsilon^{*}\cdot p_{B})\frac{A_{2}(q^2)}{m_{B}+m_{K^{*}}}\nonumber\\
&&\qquad+iq_{\mu}(\epsilon^{*}\cdot p_{B})\frac{2m_{K^{*}}}{q^2} \left(A_{3}(q^2)-A_{0}(q^2)\right)
+\varepsilon_{\mu\nu\rho\sigma}\epsilon^{*\nu}p_{B}^{\rho}p_{K^{*}}^{\sigma} \frac{2V(q^2)}{m_{B}+m_{K^{*}}}~,\\
&&\langle K(p_{K^{*}})|\bar{s}\sigma_{\mu\nu}q^{\nu}(1+\gamma_{5})b|B(p_{B})\rangle
= i\varepsilon_{\mu\nu\rho\sigma}\epsilon^{*\nu}p_{B}^{\rho}p_{K^{*}}^{\sigma}2T_{1}(q^2)
+T_{2}(q^2)\left\{\epsilon_{\mu}^{*}(m_{B}^{2}-m_{K^{*}}^{2})\right.\nonumber\\
&&\left.\qquad-(\epsilon^{*}\cdot p_{B})(p_{B}+p_{K^{*}})_{\mu}\right\}
+T_{3}(q^2)(\epsilon^{*}\cdot p_{B}))\left\{q_{\mu}-\frac{q^2}{m_{B}^{2}-m_{K^{*}}^{2}}(p_{B}+p_{K^{*}})_{\mu}\right\}~,
\end{eqnarray}
with
\begin{eqnarray}
&&\langle K^{*}|\partial_{\mu}A^{\mu}|B\rangle = 2m_{K^*}(\epsilon^{*}\cdot p_{B})A_{0}(q^2),\nonumber\\
&&A_{3}(q^2) = \frac{m_{B}+m_{K^*}}{2m_{K^*}}A_{1}(q^2) - \frac{m_{B}-m_{K^*}}{2m_{K^*}}A_{2}(q^2),\nonumber\\
&&A_{0}(0) = A_{3}(0),~~~~T_{1}(0) = T_{2}(0).
\end{eqnarray}
Using the above definition of the form factors, the decay amplitudes for $B^{+} \to K^{(*)+}\ell^{+}\ell^{-}$ corresponding to Fig.\ref{fig:BKlla} can be obtained,
\begin{eqnarray}
\mathcal{M}_{a}(B^{+} \to K^{(*)+}\ell^{+}\ell^{-}) &=&
\frac{G_{F}\alpha}{\sqrt{2}\pi}V_{ts}^{*}V_{tb}
[\bar{u}(p_{1})\gamma_{\mu}(A_{K^{(*)}} + B_{K^{(*)}}\gamma_{5})v(p_{2})
 \nonumber \\
& &+ m_{\ell}\bar{u}(p_{1})D_{K^{(*)}}\gamma_{5}v(p_{2})]
\end{eqnarray}
with
\begin{eqnarray}
&&A_{K}=p_{B}\left[C_{9}^{eff}f_{+}^{B \to K}(q^2)+2C_{7}^{eff} \frac{m_{b}}{m_{B}+m_{K}}f_{T}^{B \to K}(q^2)\right]~,\nonumber\\
&&B_{K}=p_{B}C_{10}f_{+}^{B \to K}(q^2)~,\nonumber\\
&&D_{K}=C_{10}\frac{m_{B}^{2}-m_{K}^{2}}{q^{2}} \left(f_{0}^{B \to K}(q^2)-f_{+}^{B \to K}(q^2)\right) -C_{10}f_{+}^{B \to K}(q^2) \nonumber
\end{eqnarray}
for the pseudoscalar meson $K$, and
\begin{eqnarray}
&&A_{K^*} = [A\varepsilon_{\mu\nu\rho\sigma}\epsilon^{*\nu}p_{B}^{\rho}p_{K^{*}}^{\sigma}-iB\epsilon_{\mu}^{*}+iC(\epsilon^{*}\cdot p_{B})(p_{B}+p_{K^*})_{\mu}+iD(\epsilon^{*}\cdot p_{B})(p_{B}-p_{K^*})_{\mu}]/2 ,\nonumber\\
&&B_{K^*} = [E\varepsilon_{\mu\nu\rho\sigma}\epsilon^{*\nu}p_{B}^{\rho}p_{K^{*}}^{\sigma}-iF\epsilon_{\mu}^{*}+iG(\epsilon^{*}\cdot p_{B})(p_{B}+p_{K^*})_{\mu}+iH(\epsilon^{*}\cdot p_{B})(p_{B}-p_{K^*})_{\mu}]/2 ,\nonumber\\
&&D_{K^*}=0
\end{eqnarray}
for the vector meson $K^{*}$. Here,
\begin{eqnarray}
A &=& \frac{2}{m_{B}+m_{K^*}}C_{9}^{eff}V(q^2) + \frac{4m_{b}}{q^2}C_{7}^{eff}T_{1}(q^2)~,\nonumber\\
B &=& (m_{B}+m_{K^*}) \left[C_{9}^{eff}A_{1}(q^2) + \frac{2m_{b}}{q^2}(m_{B}-m_{K^*})C_{7}^{eff}T_{2}(q^2)\right]~,\nonumber\\
C &=&\frac{1}{m_{B}^{2}-m_{K^*}^{2}}\left[(m_{B}-m_{K^*})C_{9}^{eff}A_{2}(q^2) + 2m_{b}C_{7}^{eff} \left(T_{3}(q^2) +\frac{m_{B}^{2}-m_{K^*}^{2}}{q^2}T_{2}(q^2)\right)\right]~,\nonumber\\
D &=& \frac{1}{q^2} \left[ C_{9}^{eff} \left( (m_{B}+m_{K^*})A_{1}(q^2) - (m_{B}-m_{K^*})A_{2}(q^2) - 2m_{K^*}A_{0}(q^2) \right) \right. \nonumber\\
&&\left.-2m_{b}C_{7}^{eff}T_{3}(q^2) \right],\nonumber\\
E &=& \frac{2}{m_{B}+m_{K^*}}C_{10}V(q^2)~,\nonumber\\
F &=& (m_{B}+m_{K^*})C_{10}A_{1}(q^2),\nonumber\\
G &=&\frac{1}{m_{B}+m_{K^*}}C_{10}A_{2}(q^2)~,\nonumber\\
H &=& \frac{1}{q^2}C_{10}\left[(m_{B}+m_{K^*})A_{1}(q^2) - (m_{B}-m_{K^*})A_{2}(q^2)-2m_{K^*}A_{0}(q^2)\right]~.
\end{eqnarray}

The contributions from the light and heavy neutrinos are shown in the Feynman diagrams of Fig.\ref{fig:BKllb} and Fig.\ref{fig:BKllc}, respectively. For simplification, we suppose that only one heavy Majorana neutrino exists in type-I seesaw model.
The decay amplitudes for $B^{+}\to K^{(*)+}\ell^{+}\ell^{-}$ corresponding to Fig.\ref{fig:BKllb} and Fig.\ref{fig:BKllc} can be expressed as
\begin{eqnarray}
\label{eq:Mb}
&&\mathcal{M}_{b}(B^{+}\to K^{(*)+}\ell^{+}\ell^{-}) =
-G_{F}^{2} V_{us}^{*}V_{ub} f_{B}f_{K^{(*)}} \frac{\bar{u}(p_{1})
p\!\!\!/_{B} p\!\!\!/_{\nu}  Q^{K^{(*)}} (1-\gamma_{5}) v(p_{2})}{p_{\nu}^{2}}~,\\
\label{eq:Mc}
&&\mathcal{M}_{c}(B^{+}\to K^{(*)+}\ell^{+}\ell^{-}) =
iG_{F}^{2} V_{us}^{*}V_{ub} V_{\ell N}^2 f_{B}f_{K^{(*)}} \frac{\bar{u}(p_{1})
p\!\!\!/_{B} p\!\!\!/_{N}  Q^{K^{(*)}} (1-\gamma_{5}) v(p_{2})}{p_{N}^{2} - m_{N}^{2} + i\Gamma_{N}m_{N}}~,
\end{eqnarray}
where $Q^{K}=p\!\!\!/_{K}$ and $Q^{K^{*}}=-im_{K^{*}}\varepsilon\!\!\!/^{*}$.
$f_{B}~(f_{K},~f_{K^*})$ is the decay constant of $B~(K,~K^*)$ meson. $p_{\nu}$ and $p_{N}$ stand for the four-momentum of the light neutrino $\nu$ and the heavy one $N$, respectively. $\Gamma_{N} \approx 2 \sum_{\ell} |V_{\ell N}|^2 (m_N/m_{\tau})^5 \times \Gamma_{\tau}$ represents the total decay width of the Majorana neutrino with $V_{\ell N}$ denoting the mixing matrix element between $\ell$ and $N$~\cite{Cvetic:2010rw}.
The decay rates can be written as
\begin{eqnarray}
\mathcal{B}(B^+ \!\to\! K^{(*)+}\ell^{+}\ell^{-})  =  \frac{\tau_{B}}{2m_{B}(2\pi)^{5}} \int |\mathcal{M}|^{2}  \frac{|\vec{p}_{1}^{B}|}{4m_{B}}
\frac{|\vec{p}_{2}^{*}|}{4\sqrt{s_{2K^{(*)}}}} d\Omega_{1}^{B}d\Omega_{2}^{*}ds_{2K^{(*)}}~, \label{eq:BranchingRatio}
\end{eqnarray}
with the total decay amplitude $\mathcal{M} = \mathcal{M}_{a} + \mathcal{M}_{b} + \mathcal{M}_{c}$, where $\vec{p}_{1}^{\,B}$ ($\vec{p}_{2}^{\,\,*}$) and $d\Omega_{1}^{B}$ ($d\Omega_{2}^{*}$)
denote the 3-momentum and solid angle of charged lepton $\ell^{+}$ ($\ell^{-}$)
in the rest frame of B meson ($\ell^{-}K^{(*)}$ system), respectively.
It is found that the contribution from Fig.\ref{fig:BKllb}
and interference terms is about five orders less than that from SM and can be neglected.
\begin{figure}[!ht]
\begin{center}
\subfigure[]{ \label{fig:BKll1}
\includegraphics[scale=0.7]{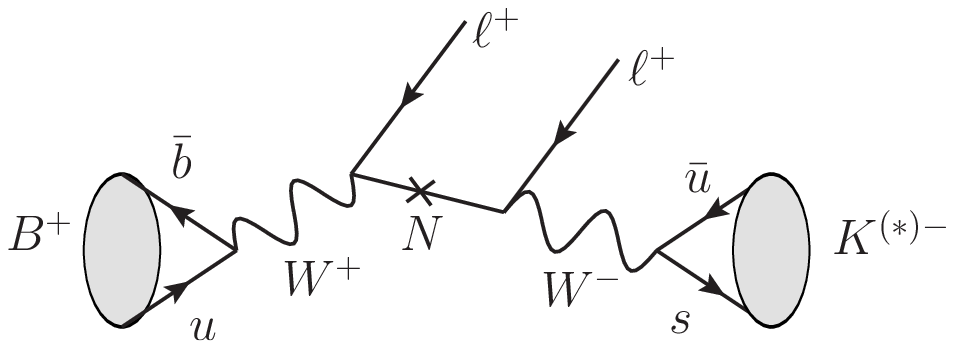}
}
\subfigure[]{ \label{fig:BKll2}
\includegraphics[scale=0.7]{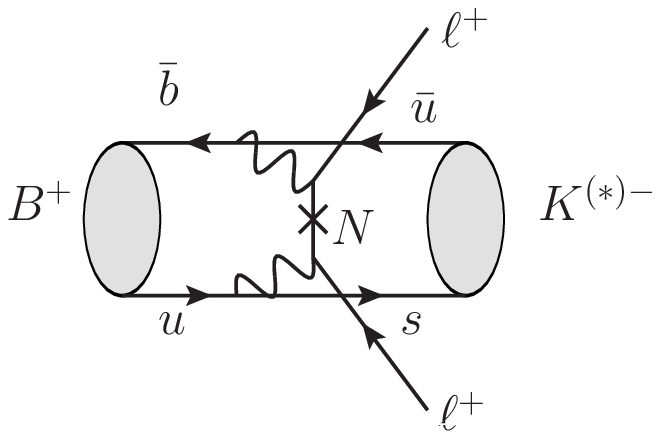}
}
\caption{Feynman diagrams for $B^{+}\to K^{(*)-}\ell^{+}\ell^{+}$ via
Majorana neutrino exchange.}\label{fig:N}
\end{center}
\end{figure}

The same-sign $\Delta L=2~$ LNV processes $B^{+}(p_B)\to K^{(*)-} (p_{K^{(*)}}) \ell^{+}(p_{1}) \ell^{+}(p_{2})$ is more sensitive to the new physics models. These decay channels may occur via Majorana neutrino exchange, especially provide evident signal if the mediate Majorana neutrino is on-shell. The dominant contribution is from the Feynman diagram of Fig.\ref{fig:BKll1},
while the contribution from Fig.\ref{fig:BKll2} is small enough to be neglected as concluded in~\cite{Bao:2012vq,Wang:2014lda}.
The corresponding decay amplitudes of $B^{+}\to K^{(*)-}\ell^{+}\ell^{+}$ are
\begin{eqnarray}
\mathcal{M}(B^{+}\to K^{(*)-}\ell^{+}\ell^{+}) &=&
-iG_{F}^{2}V_{us}^{*}V_{ub}V_{\ell N}^2 f_{B}f_{K}m_{N}   \nonumber\\
&& \left[\frac{\bar{u}(p_{1}) p\!\!\!/_{B}  Q^{K^{(*)}} (1+\gamma_{5}) v(p_{2})}
{p_{N}^{2}-m_{N}^{2}+i\Gamma_{N}m_{N}}
+\frac{ \bar{u}(p_{1})  Q^{K^{(*)}}  p\!\!\!/_{B} (1+\gamma_{5}) v(p_{2})}
{p_{N}^{\prime2}-m_{N}^{2}+i\Gamma_{N}m_{N}}\right].~~
\label{eq:MLNV}
\end{eqnarray}
Then branching ratios of these same-sign charged dilepton decays can be readily obtained by the same way as in eq.\eqref{eq:BranchingRatio}.
\section{Numerical Analysis}
The $B\to K$ form factors are parameterized by the following formulae~\cite{Wu:2009kq},
\begin{eqnarray}
&&f_{+(T)}^{B\to K}(q^2)=\frac{r_1}{1-q^2/m_{\rm{fit}}^2}
+\frac{r_2}{1-q^2/m_{B_s^*(1^-)}^2}~,\\
\label{eq:FFK+T}
&&f_{0}^{B\to K}(q^2)=\frac{r_2}{1-q^2/m_{\rm{fit}}^2}~,
\label{eq:FFK0}
\end{eqnarray}
where $m_{B_s^*(1^-)}=5.413~\mathrm{GeV}$ is the mass of the $B_s^*(1^-)$. The values of other parameters $r_{1(2)}$ and $m_{\rm{fit}}^2$ are collected in Table \ref{tab:FFK}. The parameters for $B\to K^*$ form factors are obtained from the calculation of AdS/QCD at low-to-intermediate $q^2$ and the lattice data at high $q^2$ \cite{Ahmady:2015fha}. The seven independent form factors can be expressed as the formula of
\begin{eqnarray}
F(q^2)=\frac{F(0)}{1-a\,q^2/m_B^2+b\,q^4/m_B^4}~,
\label{eq:FFKst}
\end{eqnarray}
where $F$ stands for $A_0$, $A_1$, $A_2$, $T_1$, $T_2$, $T_3$, $V$. The corresponding values of $F(0)$, $a$, $b$ are listed in Table \ref{tab:FFKst}.
\begin{table}[h]
\centering
\caption {The inputs of $r_1$, $r_2$ and $m_{fit}^2$ for $B\to K$ form factors \cite{Wu:2009kq}.
}\label{tab:FFK}
\begin{tabular}{|c|ccc|}
\hline
 ~ &$f_{+}^{B\to K}$  &$f_{T}^{B\to K}$  &$f_{0}^{B\to K}$ \\
\hline
 $r_{1}$  &$0.8182$ &$0.893$  &$--$ \\
 $r_{2}$  &$-0.4862$ &$-0.5073$  &$0.332$ \\
 $m_{\rm{fit}}^2$  &$41.61$ &$33.13$  &$61.64$ \\
\hline
\end{tabular}
\end{table}
\begin{table}[h]
\centering
\caption {The inputs of $F(0)$ (form factor at $q^2=0$), $a$ and $b$ for $B\to K^*$ form factors $A_{0}$, $A_{1}$, $A_{2}$, $T_{1}$, $T_{2}$, $T_{3}$ and $V$ \cite{Ahmady:2015fha}.
}\label{tab:FFKst}
\begin{tabular}{|c|ccccccc|}
\hline
 ~ &$A_{0}$ &$A_{1}$ &$A_{2}$
&$T_{1}$ &$T_{2}$ &$T_{3}$ &$V$ \\
\hline
 $F(0)$ &$0.243$ &$0.244$ &$0.244$
&$0.258$ &$0.239$ &$0.157$ &$0.297$ \\
 $a$ &$1.618$ &$0.586$ &$1.910$
&$1.910$ &$0.525$ &$1.147$ &$1.934$ \\
 $b$ &$0.561$ &$-0.356$ &$1.498$
&$1.082$ &$-0.459$ &$-0.114$ &$1.089$ \\
\hline
\end{tabular}
\end{table}
\begin{table}[ht]
\begin{center}
\caption{Input parameters used in our numerical calculation \cite{Patrignani:2016xqp}.}
\label{tab:BpillInPut}
\begin{tabular}{cccc}
\hline
  $\alpha$        & $1/137$
& $m_{u}$         & $2.2~\mathrm{MeV}$\\
  $G_{F}$         & $1.16637\times10^{-5}~\mathrm{GeV}^{-2}$
& $m_{s}$         & $96~\mathrm{MeV}$\\
  $\tau_{B}$      & $1.638~\mathrm{ps}$
& $m_{c}$         & $1.27~\mathrm{GeV}$\\
  $\Gamma_{\tau}$ & $2.267\times10^{-12}~\mathrm{GeV}$
& $m_{\tau}$      & $1.777~\mathrm{GeV}$\\
  $f_{K}$         & $155.6~\mathrm{MeV}$
& $m_{K}$         & $0.4937~\mathrm{GeV}$\\
  $f_{K^*}$       & $225~\mathrm{MeV}$ \cite{Ahmady:2013cva}
& $m_{K^*}$       & $0.892~\mathrm{GeV}$\\
  $f_{B}$         & $187.1~\mathrm{MeV}$
& $m_{B}$         & $5.28~\mathrm{GeV}$\\
  $\lambda$       & $0.225\pm0.0006$
& $A^{\prime}$    & $0.811\pm0.015$\\
  $\bar{\rho}$    & $0.124\pm0.024$
& $\bar{\eta}$    & $0.356\pm0.015$\\
\hline
\end{tabular}
\end{center}
\end{table}

In our numerical estimation, we only focus on dimoun decay channels considering of the relatively high muon reconstruction ability in the LHCb experiment.
The CKM matrix elements are obtained by the Wolfenstein parameterization with the values of Wolfenstein parameters ($\lambda$, $A^{\prime}$, $\bar{\rho}$, $\bar{\eta}$) and the other parameters listed in Table \ref{tab:BpillInPut}.
We work in an optimistic case with $|V_{\mu N}|^2\gg |V_{\tau N}|^2,~|V_{e N}|^2$, so the interactions are completely determined by the mixing matrix element $|V_{\mu N}|^2$ \cite{Atre:2009rg}.

\begin{figure}[!ht]
\begin{center}
\subfigure[]{ \label{fig:dBSMq2}
\includegraphics[scale=0.38]{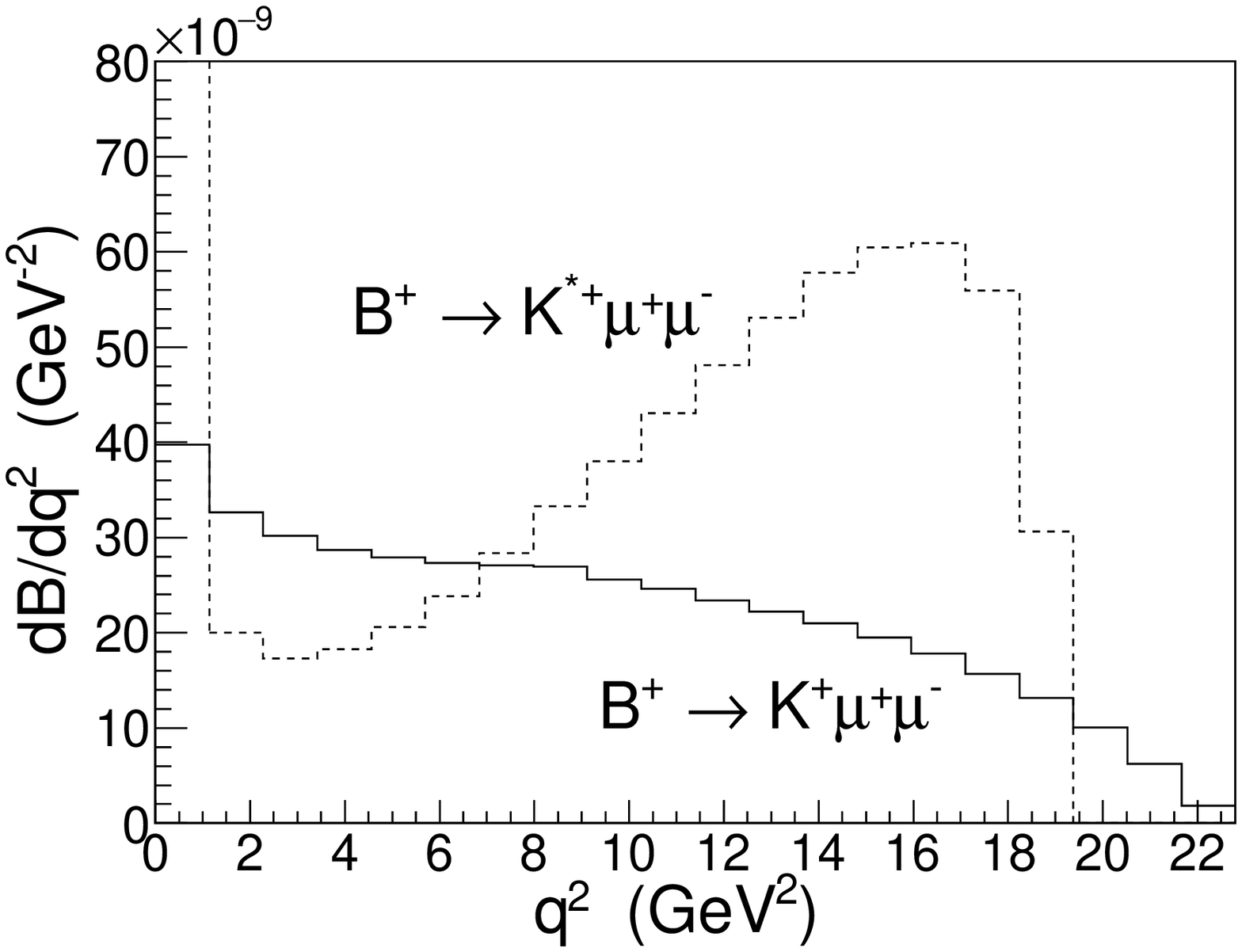}
}
\subfigure[]{ \label{fig:dBSMcosll}
\includegraphics[scale=0.38]{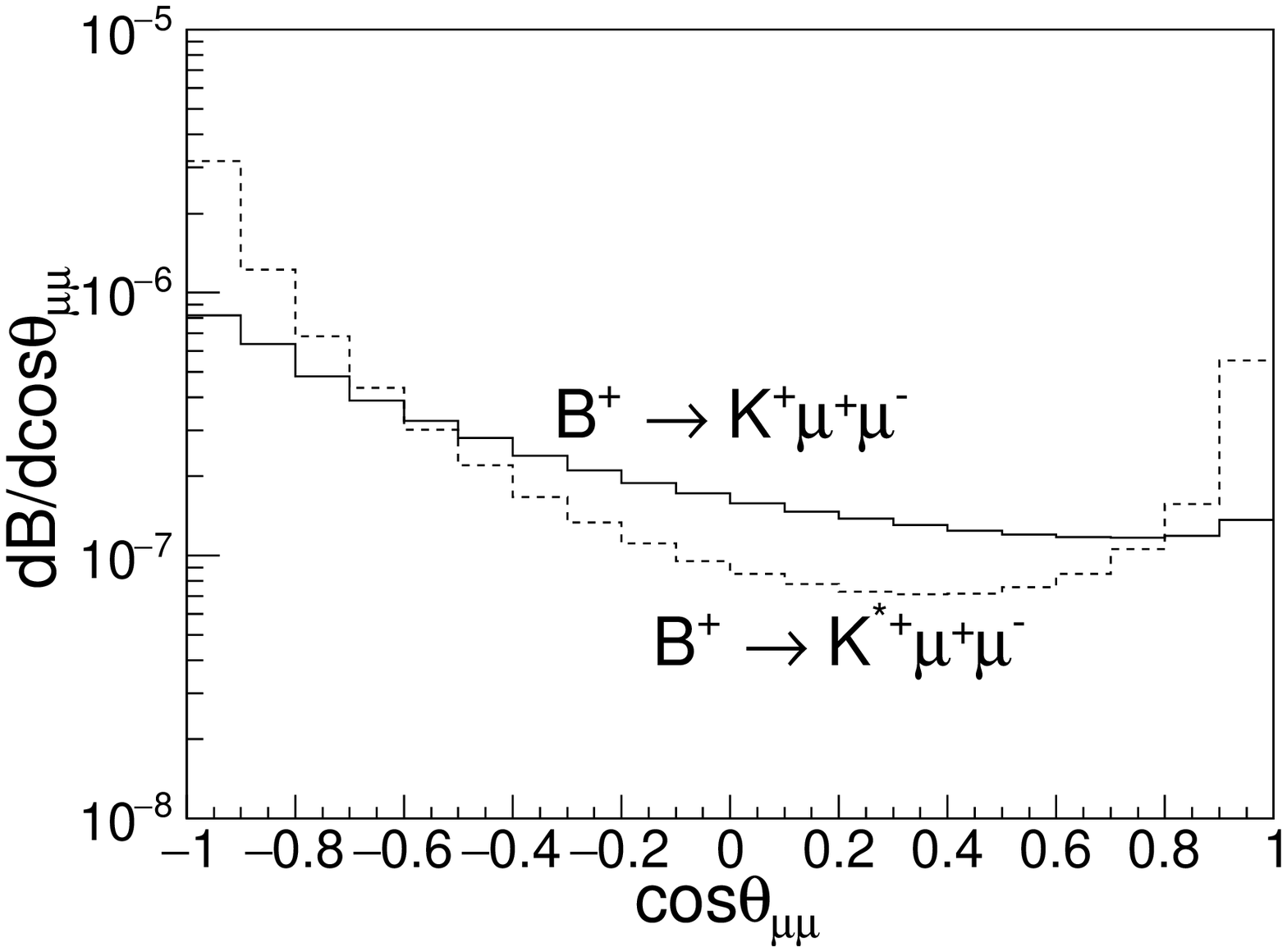}
}
\caption{(a) The invariant mass distributions and (b) the angular distributions of the opposite-sign dilepton for $\mathcal{B}(B^{+}\to K^{(*)+}\mu^{+}\mu^{-})$ in SM.
}\label{fig:dBSM}
\end{center}
\end{figure}

The SM dilepton invariant mass distributions $d\mathcal {B}(B^{+}\to K^{(*)+}\mu^+\mu^-)/dq^2$ are shown in Fig.\ref{fig:dBSMq2}. It leads to different distributions for $d\mathcal {B}(B^{+}\to K^{+}\mu^+\mu^-)/dq^2$ and $d\mathcal {B}(B^{+}\to K^{*+}\mu^+\mu^-)/dq^2$ because the form factors and the amplitudes are different for scalar and vector mesons. The integrated branching ratios are
\begin{eqnarray}
&&\mathcal {B}_{SM}(B^{+}\to K^{+}\mu^+\mu^-)=(5.21\pm0.51)\times10^{-7},\nonumber\\
&&\mathcal {B}_{SM}(B^{+}\to K^{*+}\mu^+\mu^-)=(7.63^{+0.69}_{-0.68})\times10^{-7},
\end{eqnarray}
where the dominant uncertainty comes from the CKM matrix elements and the renormalization scale variation. Experimentally, the most precise measurements on the differential branching fractions of $B^{+}\to K^{(*)+}\mu^+\mu^-$ have been performed using a data set with $3~\rm{fb}^{-1}$ of integrated luminosity collected by the LHCb detector\cite{Aaij:2014pli}. The corresponding integrated branching fractions are
\begin{eqnarray}
&&\mathcal {B}(B^{+}\to K^{+}\mu^+\mu^-)=(4.29\pm0.07(stat)\pm0.21(syst))\times10^{-7},\nonumber\\
&&\mathcal {B}(B^{+}\to K^{*+}\mu^+\mu^-)=(9.24\pm0.93(stat)\pm0.67(syst))\times10^{-7}.
\end{eqnarray}
It turns out our SM predictions for $B^{+}\to K^{(*)+}\mu^+\mu^-$ are roughly consistent with the most recent LHCb data within the range of experimental and theoretical errors, and comparable with the other SM calculation results \cite{Aliev:1997gh, Ali:1999mm, Ali:2002jg, Fu:2014uea}. The angular distributions of the opposite-sign leptons are plotted in Fig.\ref{fig:dBSMcosll} within the $B$ meson rest frame. If enough events are collected, this kind of distribution is significant to study the interactions.
\begin{figure}[!ht]
\begin{center}
\includegraphics[scale=1.0]{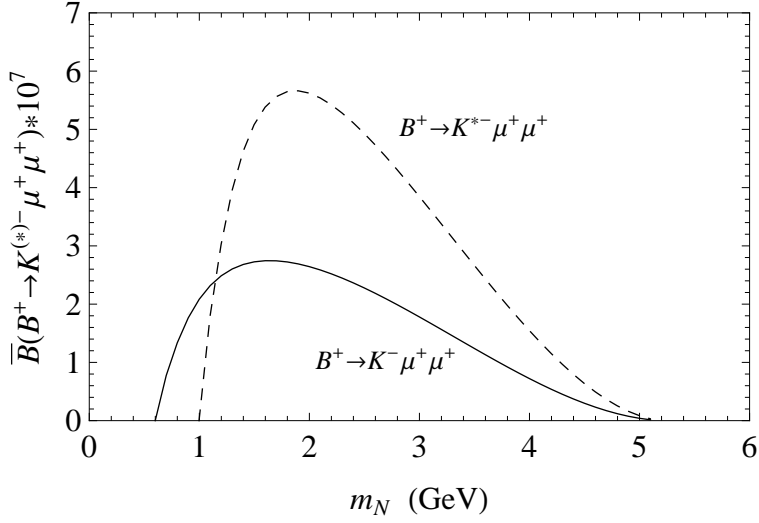}
\caption{Simplified branching ratios $\bar{\mathcal{B}}(B^{+}\to K^{(*)-} \mu^{+} \mu^{+}) = \mathcal{B} (B^{+} \to K^{(*)-} \mu^{+} \mu^{+}) / |V_{\mu N}|^2$ with respect to the Majorana neutrino mass $m_N$. }\label{fig:BKKstSSmN}
\end{center}
\end{figure}
\begin{figure}[!ht]
\begin{center}
\includegraphics[scale=1.0]{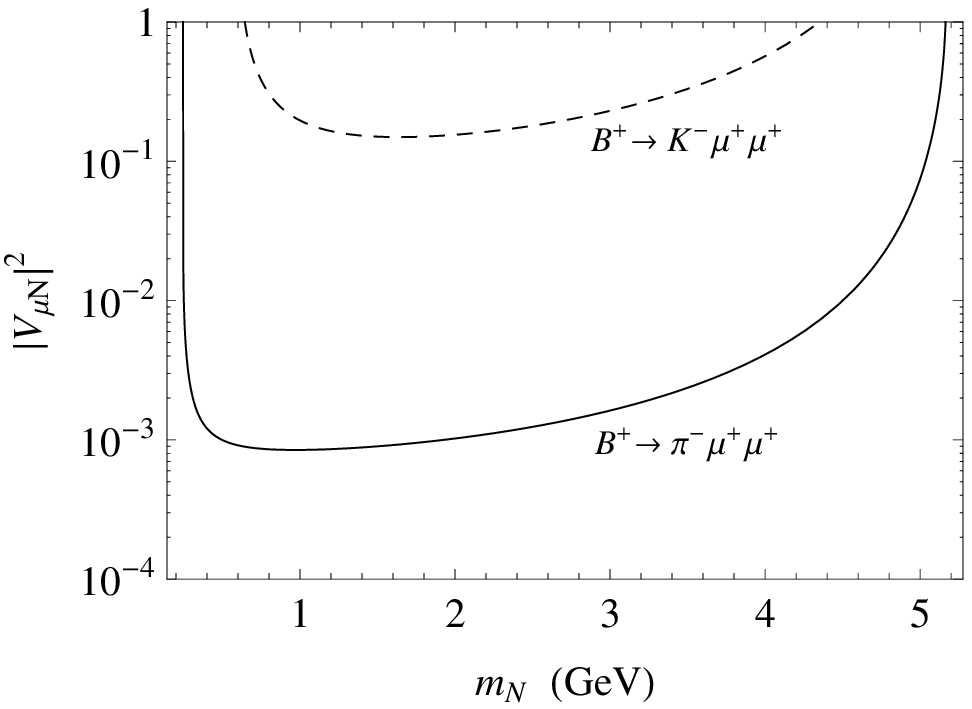}
\caption{Contour plot obtained from the experimental upper limits for LNV rare decays $B^{+}\to K^{-}\mu^{+}\mu^{+}$ and $B^{+}\to \pi^{-}\mu^{+}\mu^{+}$.
}\label{fig:KKstPi}
\end{center}
\end{figure}

The branching ratios for LNV processes depend on two new parameters, the mixing matrix element $V_{\mu N}$ and Majorana neutrino mass $m_N$.
The simplified branching ratios $\bar{\mathcal{B}}(B^{+}\to K^{(*)-} \mu^{+} \mu^{+}) = \mathcal{B} (B^{+} \to K^{(*)-} \mu^{+} \mu^{+}) / |V_{\mu N}|^2$ as functions of Majorana neutrino mass $m_N$ are plotted in Fig.\ref{fig:BKKstSSmN}. The peak is around $2~\rm{GeV}$ depending on the kinematical distributions related to the masses of $B$ and $K^{(*)}$ mesons.  Once the  Majorana neutrino mass and the mixing matrix element are fixed, the branching ratios of $B^{+}\to K^{(*)-}\mu^{+}\mu^{+}$ can be obtained. With the precise measurement on the $B$ meson decays, LHCb and BaBar collaborations have reported the upper limits for the LNV rare decays $B^{+}\to K^{(*)-}\mu^{+}\mu^{+}$~\cite{Aaij:2011ex, Lees:2013gdj},
\begin{eqnarray}
\label{eq:LHCb_LNV_BKLL}
&&\mathcal {B}(B^{+}\to K^{-}\,\mu^+\,\mu^+) <  4.1\times10^{-8}~~~\text{at $90\%$ C.L.}~~,\\
\label{eq:LHCb_LNV_BKstLL}
&&\mathcal {B}(B^{+}\to K^{*-}\mu^+\mu^+) < 5.9\times10^{-7}~~~\text{at $90\%$ C.L.}~~,
\end{eqnarray}
which can be used to constrain the parameter space between the mixing matrix element $V_{\mu N}$ and Majorana neutrino mass $m_N$. Given that the LHCb collaboration has collected the branching ratio $\mathcal{B}(B^{-}\to \pi^{+}\mu^{-}\mu^{-})\leq 4.0\times10^{-9}$ at $95\%$ C.L. \cite{Aaij:2014aba}, we plot the excluded parameter regions in the $|V_{\mu N}|^2$ versus $m_N$ plane in Fig.\ref{fig:KKstPi}. The region above the dashed (solid) line is excluded by LHCb with LNV process $B^{+} \to K^{-}(\pi^{-}) \mu^{+} \mu^{+}$ at $90\%$ ($95\%$) C.L.. As shown in this figure, the LNV rare decay channel $B^{+}\to \pi^{-} \mu^{+}\mu^{+}$ provides more rigorous constraint than the LNV $B\to K$ channel because the $B^{+}\to \pi^{-} \mu^{+}\mu^{+}$ process is much less suppressed by CKM factors than $B^{+}\to K^{(*)-}\mu^{+}\mu^{+}$ processes. However, the bounds for the new physics parameters are loosen with the  $B^{+}\to K^{*-}\mu^+\mu^+$ process until more precise experimental data are released.
\begin{table}[thb]
\begin{center}
\caption{
Majorana neutrino contributions to the branching ratios of $B^{+}\to K^{(*)-}\mu^{+}\mu^{+}$. The $m_N$ and $|V_{\mu N}|^{2}$ are the best fits corresponding to heavy quark symmetry and lattice QCD method (HQS+LQCD), perturbative QCD method (PQCD) and light cone QCD sum rule method (LCSR) referred as~\cite{Wang:2014lda}.
}
\label{tab:NPBr}
\begin{tabular}{|c|c|c|c|c|}
\hline
  ~   & $m_N$  &$|V_{\mu N}|^{2}$   & $B^{+}\to K^{-}\mu^{+}\mu^{+}$     & $B^{+}\to K^{*-}\mu^{+}\mu^{+}$\\
\hline
\multirow{3}*{HQS+LQCD}
  &~$1~\mathrm{GeV}$~  &~$8.52\times10^{-4}$~  &  $1.78\times10^{-10}$  & $1.50\times10^{-11}$\\
\cline{2-5}
  &$2~\mathrm{GeV}$  &$1.03\times10^{-3}$  &  $2.73\times10^{-10}$  & $5.79\times10^{-10}$\\
\cline{2-5}
  &$3~\mathrm{GeV}$  &$1.63\times10^{-3}$  &  $2.90\times10^{-10}$  & $6.27\times10^{-10}$\\
\hline
\multirow{3}*{PQCD}
  &$1~\mathrm{GeV}$  &$7.10\times10^{-4}$  &  $1.48\times10^{-10}$  & $1.25\times10^{-11}$\\
\cline{2-5}
  &$2~\mathrm{GeV}$  &$8.58\times10^{-4}$  &  $2.27\times10^{-10}$  & $4.82\times10^{-10}$\\
\cline{2-5}
  &$3~\mathrm{GeV}$  &$1.36\times10^{-3}$  &  $2.42\times10^{-10}$  & $5.23\times10^{-10}$\\
\hline
\multirow{3}*{LCSR}
  &$1~\mathrm{GeV}$  &$1.33\times10^{-4}$  &  $2.77\times10^{-11}$  & $2.35\times10^{-12}$\\
\cline{2-5}
  &$2~\mathrm{GeV}$  &$1.61\times10^{-4}$  &  $4.26\times10^{-11}$  & $9.05\times10^{-11}$\\
\cline{2-5}
  &$3~\mathrm{GeV}$  &$2.55\times10^{-4}$  &  $4.54\times10^{-11}$  & $9.81\times10^{-11}$\\
\hline
\end{tabular}
\end{center}
\end{table}
%

\begin{figure}[t]
\begin{center}
\subfigure[]{ \label{fig:dq2}
\includegraphics[scale=0.35]{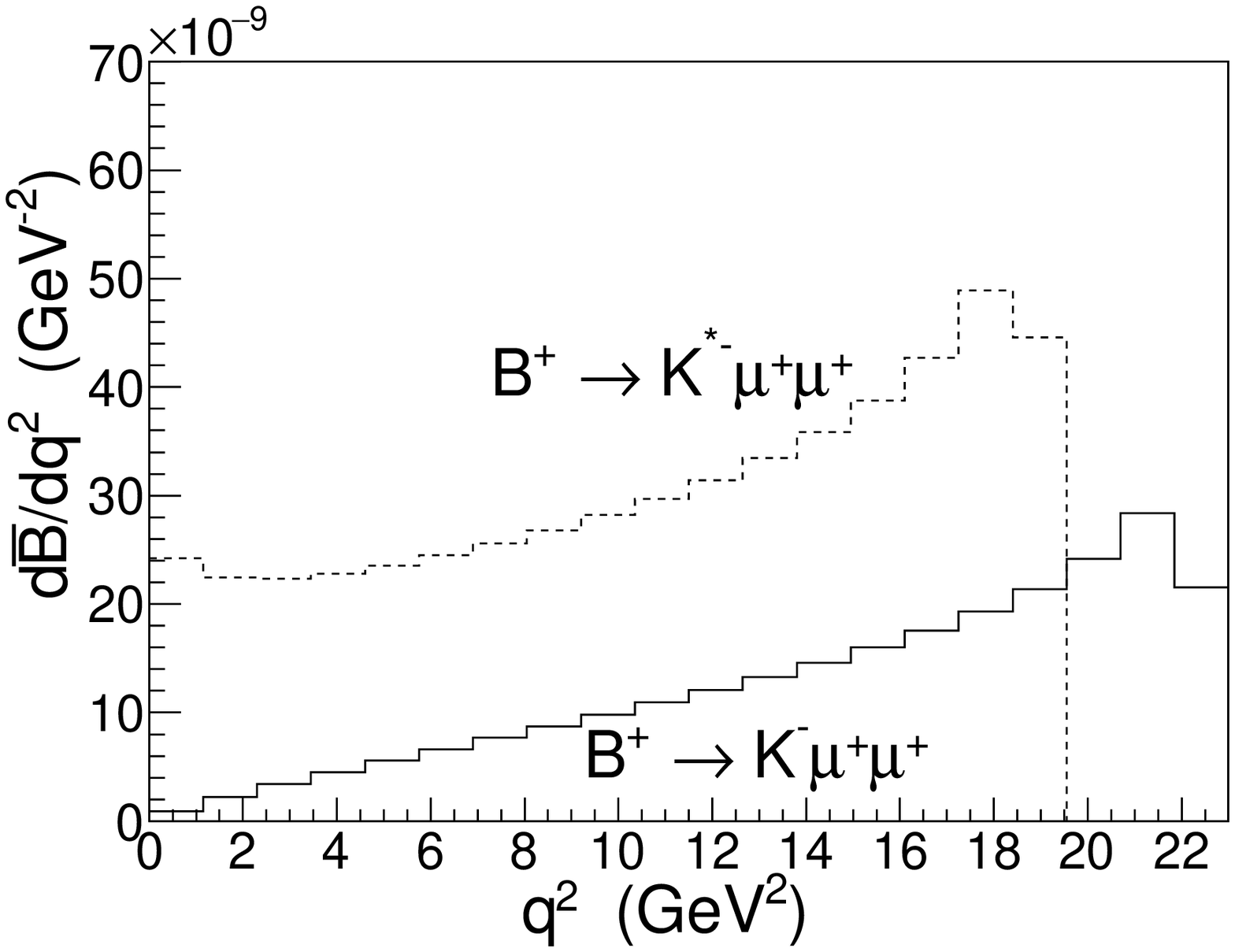}
}
\subfigure[]{ \label{fig:dcosll}
\includegraphics[scale=0.35]{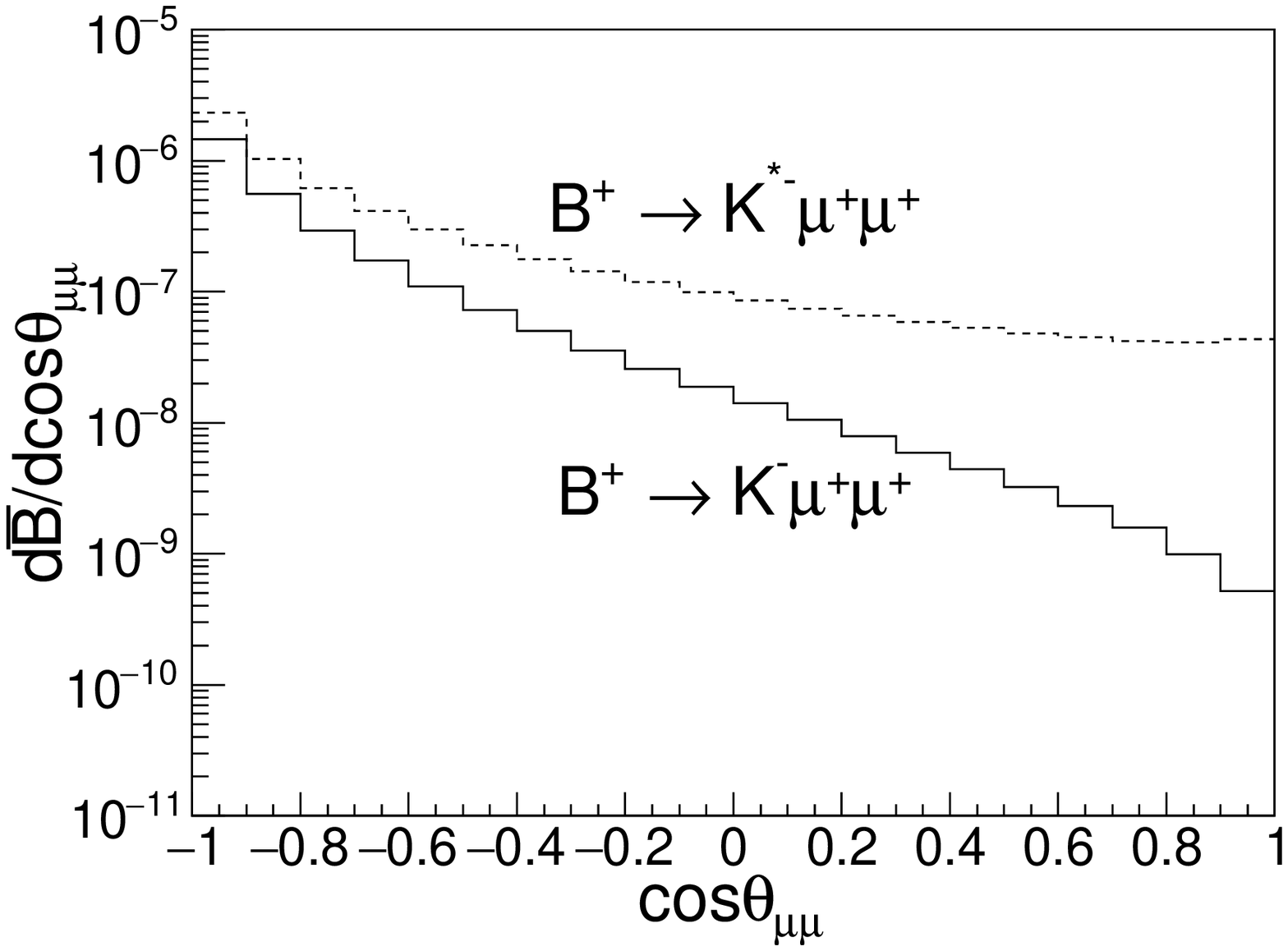}
}
\caption{(a) The invariant mass distributions and (b)the angular distributions of the same-sign dilepton for $\bar{\mathcal{B}}(B^{+}\to K^{(*)-} \mu^{+} \mu^{+})$ with $m_N=2~\rm{GeV}$.
}\label{fig:dq2dcosll}
\end{center}
\end{figure}

The best fitting values of $|V_{\mu N}|^{2}$ and $m_{N}$ obtained from our previous work \cite{Wang:2014lda} on $B^+\to \pi^-\mu^+\mu^+$ agree with the LHCb upper limit \cite{Aaij:2014aba}. As a result, we listed the branching ratios for LNV processes $B^{+}\to K^{(*)-}\mu^{+}\mu^{+}$ with the best fitting values of $|V_{\mu N}|^{2}$ and $m_{N}$ in Table \ref{tab:NPBr}.
There are three cases corresponding to the form factors with the heavy quark symmetry and lattice QCD method (HQS+LQCD), perturbative QCD method (PQCD) and light cone QCD sum rule method (LCSR) \cite{Wang:2014lda}. Here, we choose three typical values of $m_N$ in each case to estimate the branching ratios. It shows that the branching ratios of LNV processes $B^{+}\to K^{(*)-}\mu^{+}\mu^{+}$ induced by Majorana neutrino exchange are in agreement with the above LHCb (BaBar) measurements. Furthermore, it is possibly accessible at the future B-factory experiment with high integrated luminosity. The branching ratios of opposite-sign $B^{+}\to K^{(*)+}\mu^{+}\mu^{-}$ processes induced by the Majorana neutrino are the same as the same-sign processes $B^{+}\to K^{(*)-}\mu^{+}\mu^{+}$ which are a few orders less than the SM contributions. Once the LNV processes are observed, the differential distributions are necessary to be studied. In Fig.\ref{fig:dq2dcosll}, the invariant mass distributions and angular distributions of the same-sign dilepton for $\bar{\mathcal{B}}(B^{+}\to K^{(*)-} \mu^{+} \mu^{+})$ with $m_N=2~\rm{GeV}$ are presented. The same-sign process has different distributions in comparison with the opposite-sign process. These distributions can be used to investigate the decay properties of hadron and give hints to the new physics models.
\section{Summary}
LNV processes have been widely studied in the search for the new physics beyond the SM. The intermediate Majorana neutrino exchange LNV processes provide evidence for the Majorana nature of neutrinos. With the precision measurements on the B meson decays,  $\Delta L=2$ semileptonic decay processes have been elaborately calculated for the hints of new physics. We study the $B^{+} \to K^{(*)+}\mu^+\mu^-$ both in SM and type-I seesaw model. The decay branching ratios are roughly consistent with the experimental measurements.
We also investigate the LNV decays $B^{+}\to K^{(*)-}\mu^{+}\mu^+$. Parameter constraints on $m_N$ and $|V_{\mu N}|^2$, obtained with the experimental upper limits for $B^{+}\to K^{(*)-}\mu^+\mu^+$ processes, is less strict than the constraints from the $B\to \pi$ process. Thus, utilizing the best fitting for Majorana neutrino mass $m_{N}$ and the mixing matrix elements $|V_{\mu N}|^{2}$ obtained in our previous work, we give the branching ratios of the same-sign charged dilepton processes $B^{+}\to K^{(*)-}\mu^{+}\mu^+$ which is lower than the LHCb (BABAR) upper limits obtained in 2012 (2014), and the Majorana neutrino contributions to the branching ratios of $B^{+}\to K^{(*)+}\mu^+\mu^-$ are small compared with the SM contributions. Once the LNV processes is observed in the future experiment with high luminosity, the dilepton invariant mass distributions as well as the dilepton angular distributions can shed light on the properties of the new physics interactions.
\section*{Acknowledgement}
This work is supported in part by the National Natural Science Foundation of China (NSFC) under grant Nos.11635009 and 11605075 and Natural Science Foundation of Shandong Province under grant No. ZR2017JL006.
\bibliographystyle{JHEP}
\bibliography{cite}
\end{document}